\documentclass[aps,pra,twocolumn,groupedaddress]{revtex4} 

\usepackage{amsmath}  
\usepackage{amsfonts}
\usepackage{amstext}
\usepackage{latexsym, amssymb} 
\usepackage{bm}
\usepackage[linktocpage=true,colorlinks=true,citecolor=blue]{hyperref}
\DeclareMathOperator{\Tr}{Tr} 
\begin{document}
\title{Quantum theory: the role of
  microsystems and macrosystems}
\author{Ludovico \surname{Lanz}}
\author{Bassano \surname{Vacchini}}
\affiliation{Dipartimento di Fisica
dell'Universit\`a di Milano and INFN,
Sezione di Milano
\\
Via Celoria 16, I--20133, Milan, Italy}
\author{Olaf \surname{Melsheimer}}
\affiliation{Fachbereich Physik, Philipps-Universit\"at
\\
 Renthof 7, D--35032, Marburg, Germany}
\date{\today}
\begin{abstract}
   We stress the notion of statistical experiment, which is mandatory
   for quantum mechanics, and recall Ludwig's foundation of quantum
   mechanics, which provides the most general framework to deal with
   statistical experiments giving evidence for particles. In this
   approach particles appear as interaction carriers between
   preparation and registration apparatuses. We further briefly point
   out the more modern and versatile formalism of quantum theory,
   stressing the relevance of probabilistic concepts in its
   formulation.  At last we discuss the role of macrosystems, focusing
   on quantum field theory for their description and introducing for
   them objective state parameters.
\end{abstract}
\maketitle

\section{Introduction}
\label{intro}

Quantum theory is an increasingly successful theory of matter and some
typical features that have appeared paradoxical, such as EPR
correlations, are now on the way to become a technological resource in
the realm of quantum communication. There is however a fundamental
difficulty: it appears as a theory of measurements that runs into
troubles if one describes in the most naive quantum mechanical way a
measuring device at work. A precise proof of this was recently given
by Bassi and Ghirardi\cite{Bassi-decoh}, further pointing to unitary
quantum mechanical evolution as the basic flaw and proposing the GRW
modification of the Schr\"odinger equation by a universal stochastic
process\cite{Bassi-GRW-review}. Many other proposals appeared to make
quantum theory less measuring device dependent; just to mention a few
let us recall the Bohmian interpretation (see\cite{Duerr} and
literature therein for recent review and\cite{Duerr04} for present
developments), together with a new more general framework suggested by
Adler\cite{Adler}, also motivated by open problems in high energy
physics, by which stochastic modifications to the Schr\"odinger
equation can become a more natural low energy effective theory. Our
aim in this paper is to recall and refresh the, in our opinion, very
deep reformulation of foundations of quantum mechanics that was given
by Ludwig: in this approach the very concept of microsystem is
investigated and quantum theory turns out to be a naturally incomplete
theory of it, completely satisfactory at a non-relativistic level. Its
extension to general systems cannot dispense with thermodynamics and,
in our opinion, provides a natural opening to quantum field theory.
Section~\ref{sec:from-syst-micr} stresses the essential role played by
statistical experiments in quantum mechanics; in Section~\ref{micro}
the concept of microsystem as proposed by Ludwig is recalled; in
Section~\ref{modern} we briefly review the modern formulation of
quantum mechanics, which besides Ludwig's approach also arose quite
independently in other contexts; in Section~\ref{microtomacro} the
generalization to many types of microsystems is considered together
with the role played by quantum field theory in order to cope with
this situation; in Sect.~\ref{sec:conclusions-outlook} we finally
briefly summarize the contents and main message of the paper, and
discuss possible future developments.

\section{Statistical experiments}
\label{sec:from-syst-micr}

Quantum theory marked a turning point in physics since the basic
Galilean concept of reproducible experiments encountered a basic
crisis and it was necessary to weaken and clarify the notion of
\textit{reproducibility}. After Galileo, in the pre--quantum era, the
concept of reproducible experiment allowed to delimit a part of the
world ruled by physics, arriving for it to an atomistic underlying
model, the interactions between the elementary components being the
universal unifying core of the huge phenomenological complexity. 
The quantum era is characterized by the evidence that naive
reproducibility fails in experiments focused on microsystems and must
be generalized by the much subtler concept of reproducibility of
statistical experiments, so that just the most fundamental physics
becomes \textit{essentially statistical}. An experiment deals with a
system reprepared in a well fixed way for a large number of
independent experimental runs and only the frequencies of the well
defined events one is looking for in these runs are the result of the
experiment and have a counterpart in the underlying theory. What is
going on during a single run obviously belongs to reality, but not in
all aspects to physics. To physics belongs what we have described as
\textit{fixable} in the repetitions and the frequencies of well
identified changes. In suitable conditions the essentially statistical
character we have pointed out can be neglected and then the previously
mentioned classical description of matter appears and provides a
conceptual structure leading through a procedure called
\textit{quantization} to the actual quantum theory.  This could feed
the believe that \textit{quantization} of classical frameworks is
fundamental enough to catch the extremely vast phenomenology of
physical systems. On the contrary, basing on Ludwig's foundations of
quantum theory we shall take such phenomenology as the starting point
of quantum theory and as its natural completion. It is present day
technology offering high vacuum techniques, highly efficient
detectors, highly controllable sources and devices for trapping and
handling single microsystems that provides most direct evidence for
the peculiarity of microsystems.  In the more commonly observed
phenomenology what happens inside a certain space region until a
certain time influences in future times the adjacent space regions
involving simultaneously an infinity of space points. On the contrary
if an elementary microsystem is prepared in this region (e.g.  an
attenuated source is located there) a process is generated starting at
some future time point from \textit{one} single space point at an
appreciable distance (may be also an enormous distance in
astrophysical extrapolations) which then expands inside its future
light cone. In these preparations an effect is triggered only at one
space-time point in the universe even if detectors would be placed
everywhere. The position of this point and often the whole process
contained in its future light-cone is a stochastic variable; by many
runs frequencies can be established and repeating this whole procedure
the reproducibility of these frequencies can be controlled and one
gets evidence of a statistical law. This statistical law depends in
general strongly also on the devices placed between the source and the
points from which macroscopic processes can start. Still more
impressive is the most direct generalization of a microsystem
consisting of two elementary components: a source of such system
causes two processes inside the future light cones of two space-time
points, with stochasticity as before but with the restriction of
absolutely strict correlations between the two parts of the whole
processes, associated to conservation rules, quite independently of
the distance of the two points. This is a preparation such that only
at two space-time points in the universe a process begins due to
preparation, consisting of two parts, extremely strongly correlated
each other, one part showing a stochastic character. Often the
processes started by microsystems behave as further sources of
elementary and composed microsystems randomly involving a finite
number of space-time points correlated among themselves and starting
points of further correlated processes.  The fundamental consequence
of quantum theory is the statistical flavour attached to experiments,
by this refinement of the concept of reproducibility. Since
operatively a statistical experiment is a much more intriguing
enterprise, where some conditions must be explicitly selected and then
they must be controlled and guaranteed during several runs, one can
indeed expect that the mathematical representation of what is done in
an experiment must have a higher complexity as it was in pre--quantum
physics. Ludwig's work can be understood as a fundamental
justification of the new mathematical tools. In setting up an
experiment aiming to establish results in some physical context,
previous chapters of physics are taken as consolidated, some
pretheories are taken as given and already enter in the language that
is used by people setting the experiment.  So physicists, engineers,
technicians building apparatuses for a high energy experiment aiming
e.g. to verify aspects of the standard model, use experience and
knowledge coming from a well established and much simpler
phenomenology: it turns out that they have under complete control a
huge part of what goes on in the experiment, but obviously not these
aspects of reality that the experiment is challenging. In experiments
in the quantum era the technology of suitable triggering of
apparatuses with correlation and anticorrelation settings is
superposed to more classical arrangements related e.g. to Euclidean
geometry, to time determinations, to phase space of classical
mechanics and so on.

\section{The Notion of microsystem}
\label{micro}

In his axiomatic approach to the foundations of quantum mechanics
Ludwig proposed to take as fundamental domain of the theory the
statistical experiments with single microsystems and the frequencies
of the related phenomena. Instead of the particles themselves one
considers the macroscopic setup of any real experiment, which can be
divided in a preparation procedure and a registration procedure, both
to be described in terms of pretheories.  A simple example of
preparation apparatus could be an accelerator plus target, while a
typical registration apparatus could be a bubble chamber.  Once this
experimental setup is suitably described, one considers the rate
according to which microsystems prepared with the given preparation
apparatus trigger the assigned registration apparatus: these are the
frequencies to be compared with the quantum mechanical laws.  The
general scheme of a statistical experiment can be depicted as follows,
with preparation and registration apparatuses displayed as boxes
(so-called \textit{Ludwig's Kisten}), the latter acted upon by the
former by means of a directed interaction brought about by a
microsystem
 \begin{displaymath}
               \fbox{
            \hphantom{sp}
            \vrule height 20 pt depth 20 pt width 0 pt
    \mbox{$\mathrm{preparation}
    \atop\mathrm{apparatus}
    $}
            \vrule height 20 pt depth 20 pt width 0 pt
            \hphantom{sp}
            }
            \quad
{{\mathrm{directed}\atop\mathrm{interaction}}\atop \longrightarrow}
            \quad
            \fbox{
            \hphantom{sp}
            \vrule height 20 pt depth 20 pt width 0 pt
    \mbox{$
    \textrm{registration}
    \atop
    \textrm{apparatus}
    $}
            \vrule height 20 pt depth 20 pt width 0 pt
            \hphantom{sp}
            }     \ .
\end{displaymath}
In this spirit we want to introduce the notion of microsystem as of
something which has been prepared by a preparation apparatus and
registered by a registration apparatus. To do this we need a
statistical theory, in terms of which the general structures of
preparation and registration, which can be applied both to microsystem
and macrosystem, can be described.  We will almost verbatim follow the
introductory reference\cite{Ludwig-Grundlagen} when recalling the
basic axioms, but we shall proceed in a less detailed way with respect
to the full axiomatic approach initiated in the sixties and described
in\cite{Ludwig-Foundations}.

\subsection{Statistical selection procedures}
\label{sec:stat-select-proc}

Let $M$ be the set having as elements the representatives
of the physical features whose statistics we want to
describe (in the present case we shall concentrate on microsystems).
The statistics is related to selection procedures, by which
special features may be selected. A selection procedure is
to be described by a subset $a\subset M$, corresponding to the
subset of features (microsystems) that satisfy the given
selection procedure.
We define as \textsc{ Selection Procedure} the
following mathematical structure: a set $M$ and a subset
${\cal S} \subset {P}(M)$ (where ${P}(M)$ denotes the power set of $M$) such that
\begin{description}
        \item[S~1.1]
                $
                a, b \in {\cal S},  a\subset b \Rightarrow
                b \backslash a \in {\cal S}
                $
        \item[S~1.2]
                $
                a, b \in {\cal S} \Rightarrow
                a \cap b \in {\cal S}.
                $
\end{description}
We call selection procedure both ${\cal S}$ and an element $a$ of
${\cal S}$.  {\bf S~1.2} says that the selection procedure consisting
in selecting both according to $a$ and $b$ exists.  If $a \subset b$
we say that $a$ is {\em finer} than $b$.  {\bf S~1.1} says that if we
use two selection procedures $a$ and $b$, where $a$ is finer than $b$,
the rest of the objects $x \in b$, which do not satisfy the finer
criterion $a$, still constitute a selection procedure.  Note that in
this construction it is not necessarily $M\in{\cal S}$.  Only if some
selection has been done on its elements $M$ acquires a physical
meaning, therefore we shall not assume that $M$ itself belongs to
$\mathcal{S}$.  In fact $a\in {\cal S}$, $M\in{\cal S}$ would lead to
$M\backslash a\in{\cal S}$ contrary to physical meaningfulness.  Let
us consider in fact a given source for a certain type of particles.
For the prepared particles we can make important assertions about the
experiments for which the particles are used, but we cannot make any
definite statement for all other particles of the same type not
prepared from this source. Thus it is meaningful not to require that
$M$ be a selection procedure.  Let us note that ${\cal S}(a)\equiv \{
b \in {\cal S} | b \subset a \} $ is a Boolean ring, while ${\cal S}$
need not even be a lattice, because $a, b \in{\cal S}$ does not
automatically imply $a \cup b \in {\cal S}$.  In particular two
selection procedures $a, b \in {\cal S}$ are called
\textsc{Coexistent} relative to $c$ if both $a \subset c$ and $b
\subset c$.

It often happens in applications that two selection
procedures $a$ and $b$, where $b$ is finer than $a$,
are not statistically independent.
Consider for example an experiment in which of the $N$ systems
prepared according to the selection procedure $a$, $N_1$
also satisfy the selection criterion of $b$: we say that
$N_1/N$ is the relative frequency of $b$ relative to $a$.
If this frequency shows to be reproducible and it is
confirmed by experiments with great number of systems, we say that $b$
is statistically dependent from $a$.
Let ${\cal
S} \subset {P}(M)$ be a selection procedure, for which
{\bf S~1.1} and {\bf S~1.2} hold, and let ${\cal T} \equiv
\{
(a,b) | a, b \in {\cal S}, b \subset a, a \ne \emptyset
\}
$: we say that ${\cal S}$ is a  \textsc{Statistical
Selection Procedure} whenever a real function $\lambda(a,b)$
with $0 \leq \lambda(a,b) \leq 1$   is defined on ${\cal T}$
such that
        \begin{description}
        \item[S~2.1]
                $
                a_1, a_2 \in {\cal S},
                a_1 \cap a_2 = \emptyset,
                a_1 \cup a_2 \in {\cal S}
                \Rightarrow
                \lambda(a_1 \cup a_2,a_1) + \lambda(a_1 \cup
                a_2,a_2)=1
                $
        \item[S~2.2]
                $
                a_1, a_2, a_3 \in {\cal S},
                a_1 \supset a_2 \supset a_3,
                a_2 \ne \emptyset
                \Rightarrow
                \lambda(a_1,a_3)=  \lambda(a_1,a_2)\lambda(a_2,a_3)
                $
        \item[S~2.3]
                $
                a_1, a_2 \in {\cal S},   a_1 \supset a_2,
                a_2 \ne \emptyset
                \Rightarrow
                \lambda(a_1,a_2) \ne 0
                $                     .
        \end{description}
$\lambda(a,b)$ is usually called the {\em conditional probability}
of $b$ relative to $a$ and represents the frequency with
which systems selected by $a$ also satisfy $b$.
If $a_1 \cup a_2$ is a selection procedure, both $a_1$ and
$a_2$ are finer than $a_1 \cup a_2$; if $a_1 \cap a_2 =
\emptyset $
they exclude each other. If $N$ systems are selected
according to $a_1 \cup a_2$, of which $N_1$ also satisfy
$a_1$ and $N_2$ satisfy $a_2$, because of $a_1 \cap a_2 =
\emptyset $
we have $N_1 + N_2 = N$: this explains {\bf S~2.1}.
If for three  selection procedures we have $a_1 \supset a_2 \supset a_3$
and $N_1$ systems are selected according to $a_1$, between
these $N_2$ according to $a_2$, between these again $N_3$
according to $a_3$, we simply have
$N_3 / N_1 = (N_2 / N_1) (N_3 / N_2)$, that is to say {\bf
S~2.2}.
If $a_1 \supset a_2 \ne \emptyset$, of the $N$ systems
chosen according to $a_1$ certainly finitely many will  also
satisfy $a_2$, which is {\bf S~3.3}.
From these axioms follows $\lambda(a_1,0)=0$ and
$\lambda(a_1,a_1)=1$; moreover, if $a_2 \cap a_3 = \emptyset$, $a_2,
a_3 \subset a_1$, we have
$
\lambda(a_1, a_2 \cup a_3) = \lambda(a_1,a_2) +
\lambda(a_1,a_3)
$.

Note that $\mu(b)=\lambda(a,b)$ is an additive measure
on the Boolean ring ${\cal S}(a)$ and for $a \supset a_1
\supset a_2$ we have $\lambda(a_1,a_2)=\mu(a_2) / \mu(a_1)$.
On the Boolean ring ${\cal S}(a)$ one can therefore recover all the
conditional probabilities $\lambda(a,b)$ from the
probability function $\mu(b)$. 

Such structure is very general and only the additional criteria by
which the family $\mathcal{S}$ is selected out from $M$ allows to
recognize relationship with physical procedures. These procedures come
from phenomenology, from known sectors of physics and technology. They
can be implemented in laboratories since appropriate language and
techniques have been developed and it is rather obvious that
apparatuses used in experimental settings are related to the whole
technical evolution by which materials were produced, that can be
sorted and adequately transformed. General thermodynamical concepts
such as local equilibrium are immediately of relevance, non
equilibrium being producible by putting different components together;
basic physical indexes, such as temperature, were recognized and led
to the feasibility of increasingly sophisticated selections.  Families
$\mathcal{S}$ of subsets satisfying only the requirements
\textbf{S~1}, which we have simply called selection procedures,
acquire the fundamental properties \textbf{S~2} only if an adequate
degree of selection has been attained, generally including some
suitable isolation or shielding device. Then when a sufficiently
selected subset $a \in \mathcal{S}$ has been obtained, further
partitions of $a$ into disjoint subsets show the statistical
regularity expressed by \textbf{S~2}, so that $\mathcal{S}$ is a
statistical selection procedure and physics can start to explain the
probability function $\lambda(a,b)$, $b\subset a$. Condition
\textbf{S~1.1} means that once we are able to perform selection $a$
and selection $b$, it is possible to build an equipment which produces
selections $a$ and $b$ together, i.e. we are considering only
compatible selection procedures: this is the practical way to produce,
starting with two selection procedures $a$ and $b$, another one $a
\cap b$ finer then $a$ and $b$ since $a \cap b\subset a$, $a \cap
b\subset b$. Phenomenology used in setting experiments seems to
satisfy this very simple and general criterion; this is often but
misleadingly described as a \textit{classical } character of
macroscopic world. Taking the concepts of time and space as already
established, associating selection procedures to space-time regions
one is lead, by relativistic causality, to assume that selection
procedures associated with two space time regions at space-like
separation from each other are compatible.  It is a selection
procedure to prepare a physical system during a certain initial time
interval inside a finite space region, then finer selections can be
done over a longer time interval and if these are statistical
selection procedures a very general statistical description of
dynamics is achieved, i.e. control of the system in the initial time
interval is often enough to allow a statistical regularity during the
time evolution of the system.

We now introduce a mathematical expression for the notion of
experimental mixture. Considering a selection procedure ${\cal S}$, a
partition of $a\in{\cal S}$ of the form $a=\cup_{i=1}^{n} b_i$, with
$b_i \in {\cal S}$ and mutually disjoint is called a
\textsc{Decomposition} of $a$ in the $b_i$, and $a$ is called a
\textsc{Mixture} of the $b_i$. Since the set ${\cal S}(a)$ is a
Boolean ring a decomposition of $a$ is simply a disjoint partition of
the unit element $a$ of ${\cal S}(a)$. With the above defined additive
measure $\mu(b)$ over ${\cal S}(a)$ we have $\sum_{i=1}^{n}
\mu(b_i)=1$, $\mu(b_i)=\lambda(a,b_i)$ being the weights of $b_i$ in
$a$. If we experimentally choose $N$ systems according to $a$, and of
these $N_i$ are further selected according to $b_i$, the relations
$N_i / N \approx \mu(b_i)$ must be verified in physical approximation.
This should however not induce the reader to confuse the notion of
selection procedure with that of ensemble, which will be introduced
later on.

\subsection{Preparation and registration}
\label{sec:prep-registr}

Exploiting the above defined notions of selection procedure
and of statistical selection procedure we want to introduce on
$M$ (which is expected to become the set of microsystems)
suitable mathematical structures, so as to interpret its
elements as physical systems, in that they can be prepared
and registered. Let a structure ${\cal Q}\subset{P}(M)$
be given on $M$, which we call \textsc{ Preparation
Procedure}, such that ({\bf A} standing for axiom)
        \begin{description}
        \item[A~1] ${\cal Q}$ is a statistical selection procedure.
        \end{description}
The elements of ${\cal Q}$ are representatives of
well-defined technical processes, to be described by
pretheories and not by  quantum mechanics itself,
thanks to which  microsystems can be produced in large
numbers. The mathematical relation $x\in a$ ($a \in {\cal
Q}$) means: $x$ has been obtained according to the
preparation  procedure $a$. There are very many examples of
preparation  procedures, e.g., an ion-accelerator together with
the apparatus which generates the ion-beam.
We denote by $\lambda_{{\cal Q}}(a,b)$ the probability
function defined over ${\cal Q}$.
We now consider a specific physical example, in order to
make this construction clearer. We take an experimental
apparatus which generates couples (1,2) of spin 1/2 particles with
total spin 0  and emits them in opposite directions. As
preparation  procedure for the system 1 we consider the
apparatus consisting of the  preparation  apparatus for the
couple (1,2) and an  apparatus detecting the $z$ component of
the spin of system 2. This apparatus
gives us three different preparation  procedures for the
system 1.  Preparation procedure $a_1^3$: all prepared
systems 1 independent of the detection on system 2;
preparation  procedure $a_1^{3+}$: all systems 1, by which a
positive $z$ component has been detected for system 2;
preparation  procedure $a_1^{3-}$: all systems 1, by which a
negative $z$ component has been detected for system 2.
We obviously have
$a_1^{3+} \subset a_1^3$,
$a_1^{3-} \subset a_1^3$,
$a_1^{3+} \cap a_1^{3-}= \emptyset$ and
$a_1^3=a_1^{3-} \cup a_1^{3+}$ represents a decomposition
of $a_1^3$. In this particular case the weights are given by
$\mu(a_1^{3\pm})=\lambda(a_1^3,a_1^{3\pm})= {1/2}$.

We now want to introduce the notion of  registration. Let
there be on $M$ two further structures, the set of \textsc{ Registration
Procedures} ${\cal R}\subset {P}(M)$ and the set of
\textsc{ Registration Methods}  ${\cal R}_0 \subset {P}(M)$, satisfying
        \begin{description}
        \item[A~2\hphantom{.1}] ${\cal R}$ is a selection procedure
        \item[A~3\hphantom{.1}] ${\cal R}_0$ is a
                                  statistical selection procedure
        \item[A~4.1] ${\cal R}_0 \subset {\cal R}$
        \item[A~4.2] From $b \in {\cal R}$ and ${\cal
            R}_0 \ni b_0 \subset b$ follows $b\in {\cal R}_0$
\item[A~4.3] To each $b \in {\cal R}$ there exists
                       a $b_0 \in {\cal R}_0$ for which
                       $b\subset b_0$.
\end{description}
These two structures correspond to the two steps of a typical
registration process: the construction and utilization of the
registration apparatus and the selection according to the changes
which have occurred or not occurred in the registration apparatus. Let
us consider for example a proportional counter: $b_0 \in {\cal R}_0$
is the set of all microsystems which have been applied to the counter;
the elements of ${\cal R}_0$ characterize therefore the construction
of the registration apparatus and its application to microsystems. For
a particular microsystem $x\in b_0$ the counter may or may not
respond: let $b_+$ (with $b_+ \subset b_0$) be the selection procedure
of all $x\in b_0$ for which the counter has responded and $b_-$ the
set of all $x\in b_0$ for which the counter has not responded. $b_+$
and $b_-$ are elements of ${\cal R}$.  {\bf A~3} accounts for the fact
that the apparatus, apart from triggering by microsystems, is a
macrosystem with statistically reproducible features. It is instead
extremely important that we do not require ${\cal R}$ to be a
statistical selection procedure. To understand this point let us come
back to the previous example. The counter characterized by $b_0$ may
respond or not, so that $b_0$ is decomposed in the two sets $b_+$ and
$b_-$, such that $b_0 = b_+ \cup b_-$ and $b_+ \cap b_- = \emptyset$.
There is however in nature no reproducible frequency $\lambda_{{\cal
    R}}(b_0,b_+)$; in fact if in a real experiment $N$ microsystems
$x_1, x_2, \ldots , x_N$ are applied to the counter, i.e., $x_1 \in
b_0, x_2 \in b_0, \ldots , x_N \in b_0$, and for $N_+$ of these the
counter has responded, the frequency $N_+ / N$ depends in an essential
way on the previous history of the microsystems, it cannot be
reproduced on the basis of the registration procedure alone.

Let us call ${\cal S}$ the smallest set of selection
procedures containing all $a\cap b$ with $a\in{\cal Q}$ and
$b\in{\cal R}$ (remember that $a\cap b$ is the set of all
microsystems that have been prepared according to $a$ and
registered according to $b$). We have ${\cal S}\subset{P}(M)$, but in the general case neither
${\cal Q}\subset{\cal S}$ nor
${\cal R}\subset{\cal S}$ will be true.
We now come to a most important statement, according to
which  preparation {\em and} registration  procedures
together give reproducible frequencies
        \begin{description}
        \item[A~5] ${\cal S}$ is a statistical selection procedure.
        \end{description}
Of course there will be some relations between the
statistics in ${\cal S}$ and those in ${\cal Q}$ and ${\cal
R}_0$.
We now want to express the fact that  preparation
procedures and  registration methods are
independent of each other; denoting with $\lambda_{{\cal
S}}(c,c')$ the probability function in ${\cal S}$ we have
        \begin{description}
        \item[A~6.1] If $a,a'\in {\cal Q}$, $a'\subset a$
                       and $b_0\in{\cal R}_0$ then
                       $\lambda_{\cal S}(a\cap b_0,a'\cap b_0)
                       =\lambda_{\cal Q}(a,a')$
        \item[A~6.2] If $a\in {\cal Q}$
                       and $b_0,b'_0 \in{\cal R}_0$, $b'_0\subset b_0$,
                       then $\lambda_{\cal S}(a\cap b_0, a\cap b'_0)=
                       \lambda_{{\cal R}_0}
                       (b_0,b'_0)$.
\end{description}
On the contrary in general $\lambda_{\cal S}(a\cap b,a'\cap b)
                       \not =\lambda_{\cal Q}(a,a')$, where 
$\lambda_{\cal Q}(a,a')$ is the frequency with which
microsystems prepared according to $a$ satisfy the finer
selection $a'$. 
\textbf{A~6.1} and \textbf{A~6.2} mean that, except for the
microsystem, the preparation and registration apparatuses do not interact.
Thus {\bf A~6.1} and {\bf A~6.2} express the directedness of the
interaction of the preparation on the registration
apparatus.

A set $M$ with three structures ${\cal Q}\subset{P}(M)$, ${\cal
  R}\subset{P}(M)$, ${\cal R}_0 \subset{P}(M)$, satisfying
{\bf A~1} to {\bf A~6} is a set of physical systems selected by a
measuring process.  As stressed at the beginning of this section the
structures we have used to introduce the notion of physical system are
not restricted to the case of microsystems, they can describe
measurements on macroscopic systems as well. Thanks to the axioms {\bf
  A~1} to {\bf A~6}, implying the independence of the preparation
procedure with respect to the registration procedure, the facts that
we have called physical systems have some reality beyond that of the
direct interpretation in terms of preparation and registration
procedures.  Intuitively this means that in the preparation
\textit{something} is produced which can be afterwards detected by the
registration apparatus. Nevertheless the physical systems that we have
introduced are still closely related to the associated production and
detection methods; it does not seem that they can be described in
terms of the objective properties that we are accustomed to ascribe to
physical systems. Speaking of self-existing objects which do not
suffer or exert any influence on the rest of the world would be
physically meaningless and, from a logical point of view,
self-contradictory.  Nevertheless in physics one seeks to describe
portions of the world as if they were isolated, in the sense that on a
given description level their interactions with the rest of the world
may be neglected. To the extent that this is possible one may
attribute objective properties to the considered system. The
introduced scheme is so far very general, being applicable both to
macrosystems and microsystems: the selection procedures in ${\cal S}$
describe a conventional \textit{classical statistics}, not exhibiting the
\textit{typical} quantum mechanical structure. The transition to quantum
statistics will be made only later with axiom {\bf QM}, thus coming to
the notion of microsystem.

\subsection{Equivalence classes}
\label{sec:equivalence-classes}

From
{\bf S~2} and
{\bf A~6} one can prove that the probability function
$\lambda_{\cal S}(c,c')$ is uniquely determined by
$\lambda_{\cal Q}$ and by the special values
        \begin{equation}
        \label{2.1}
        \lambda_{\cal S}(a\cap b_0,a\cap b)     ,
        \end{equation}
with $a\in {\cal Q}$,
$b\in{\cal R}$, $b_0\in{\cal R}_0$ and $b\subset b_0$.
$\lambda_{\cal S}(a\cap b_0,a\cap b)$ gives the frequencies
with which  microsystems prepared by $a$ and
applied to the apparatus characterized by $b_0$ trigger it
according to $b$.
The values (\ref{2.1}) are just the values
the experimental physicist obtains to compare with the theory: $N$
systems are prepared according to the  preparation
procedure $a$ and applied to the  registration method
specified by $b_0$, then one counts the number $N_+$ of
microsystems which trigger the registration  apparatus in a
definite way, specified by $b$. Within physical
approximations the number $N_+ / N$ should agree with
(\ref{2.1}): the whole statistics of experiments with
microsystems is contained in (\ref{2.1}).

To proceed further let us introduce the set ${\cal F}$ of
\textsc{Effect Processes}: ${\cal F}\equiv \{ {(b_0,b)} | b_0\in{\cal
  R}_0, b_0\ne\emptyset , b\in{\cal R},b\subset b_0 \}$.  A couple
${(b_0,b)}$ in ${\cal F}$ exactly describes the experimental situation
corresponding to the generation of an effect.  We may now write in a
simpler way the function (\ref{2.1}): denoting by $g={(b_0,b)}$ a
couple in ${\cal F}$ we define $\lambda_{\cal S}(a\cap b_0,a\cap
b)=\mu (a,g)$, where the function $\mu(a,g)$ is defined on the whole
${\cal Q} \times {\cal F}$ .  According to $\mu (a_1,g)=\mu(a_2,g)$
for all $g\in{\cal F}$ an equivalence relation $a_1 \sim a_2$ is
defined on ${\cal Q}$, which allows to partition it into equivalence
classes.  We call ${\cal K}$ the set of all equivalence classes in
${\cal Q}$: an element of ${\cal K}$ is called \textsc{Ensemble} (or
{\em state}) and ${\cal K}$ is the set of ensembles. Let us stress the
fact that an ensemble $w\in{\cal K}$ is not a subset of $M$, that is
to say, an ensemble $w$ is not a set of prepared microsystems: it is a
class of sets $a$ of prepared microsystems. The difference between
ensembles and preparation procedures is very important.  Analogously
to what has been done in ${\cal Q}$, one can introduce an equivalence
relation in ${\cal F}$: $g_1\sim g_2$ whenever $ \mu(a,g_1)=
\mu(a,g_2) $ for all $a\in{\cal Q}$. We denote by ${\cal L}$ the set
of all equivalence classes in ${\cal F}$: an element $f\in{\cal L}$ is
called \textsc{Effect} and ${\cal L}$ is the set of all effects.  Once
again one should not confuse effects and effect processes.  Through
${\tilde \mu}(w,f)=\mu(a,g)$ for $w\in{\cal K}$, $f\in{\cal L}$ and
$a\in w$, $g\in f$ a function ${\tilde \mu}(w,f)$ is defined on ${\cal
  K}\times{\cal L}$ (in the following we will simply write $\mu$
instead of $\tilde\mu$).  For the real function ${\mu}(w,f)$ on ${\cal
  K}\times{\cal L}$ we have:
        \begin{enumerate}
        \item   $0\leq \mu(w,f)\leq 1$,
        \item   $\mu(w_1,f)=\mu(w_2,f)\ \forall f\in{\cal L}
                \Rightarrow w_1=w_2$,
        \item   $\mu(w,f_1)=\mu(w,f_2)\ \forall w\in{\cal K}
                \Rightarrow f_1=f_2$,
        \item   $\exists !\ f_0\in{\cal L}$ (also
                denoted by 0) such that
                $\mu(w,f_0)=0\ \forall w\in{\cal K}$,
        \item   $\exists !\  f_1\in{\cal L}$ (also
                denoted by \openone) such that
                $\mu(w,f_1)=1\  \forall w\in{\cal K}$.
        \end{enumerate}

Mixtures on $\mathcal{Q}$ are transferred on $\mathcal{K}$, as one can
show taking into account $\lambda_{\cal S}(a\cap b_0, a'\cap b_0)=\lambda_{\cal Q}(a,a')$: let $a=\cup_{i=1}^{n}a_i$,
$a_i\in\mathcal{Q}$, $a_i\not\in \emptyset$, $a_i\cap a_j=\emptyset$ $i\ne
j$ then if $a\in w$, $a_i\in w_i$ one has
\begin{displaymath}
   w=\sum_{i=1}^{n} \lambda_{\cal Q}(a,a_i)w_i.
\end{displaymath}
By this fundamental statistical property a preparation procedure
$a\in\mathcal{Q}$ of a microsystem becomes very close to an element
$w\in \mathcal{K}$. It is however very important to be aware of the
fact that the passage from $\mathcal{Q}$ to $\mathcal{K}$ is a step by
which new mathematical entities are introduced which have a basic role
in describing the physics of a microsystem under \textit{all possible
  preparation and detection procedures}: then $w$ does not simply
represent one concrete preparation procedure of a microsystem.
By the introduction of equivalence classes some universality character
of $\mu (w,f)$ has been introduced: in the case of a microsystem these
equivalence classes contain a huge number of elements.
Actually something of the particular experimental situation described
by $a\in\mathcal{Q}$ goes lost when the equivalence class $w$ to which
$a$ belongs is considered and the inverse passage from $w$ to $a$
cannot be done if one only relies on quantum theory of microsystems.
Some paradox in quantum mechanics, e.g. EPR paradox have their roots
just in neglecting this fact, typically in an equivalence class $w\in
K$ two preparation procedures $a$ and $a' \in\mathcal{Q}$ can be
contained which are \textit{incompatible}: $a\cap a'=\emptyset$, i.e.
the two concrete selection procedures cannot be performed together. In
Ludwig's point of view the debated question of completeness of quantum
mechanics is not an issue from the very beginning.

The introduced partitions into equivalence classes of the sets ${\cal
  Q}$ and ${\cal F}$ are most important. These partitions do not
simply amount to make the theory of the considered physical systems
independent from inessential features in the construction of the
apparatuses $a\in{\cal Q}$ and $b_0\in{\cal R}_0$.  They have a much
deeper significance with regard to the physical theory. For example
the partition of ${\cal Q}$ depends in an essentially way on which and
how many effect processes are physically realizable. Restricting the
set ${\cal F}$ to a subset $\widetilde{{\cal F}}$ could imply a
coarser partition of ${\cal Q}$.  Axioms about the extension of the
sets ${\cal Q}$ and ${\cal F}$ amount to specify the theory one is
dealing with, thus indirectly identifying the described physical
systems and the possible realizable experiments.

\subsection{Quantum Mechanics}
\label{ro and f in mq}

So far we have introduced the quantities that connect
theory and experiment, that is to say the elements of
${\cal Q}$, ${\cal R}$, ${\cal R}_0$ and the functions
$\lambda_{\cal Q}$,
$\lambda_{{\cal R}_0}$,
$\lambda_{\cal S}$.
Note that contrary to the usual formulations of quantum
mechanics, neither the statistical operators (or in
particular the pure states) nor the self-adjoint operators
(describing the so-called observables) will be used for
direct comparison with experiment: the relationship between
mathematical description and experiment exclusively rests
upon the  preparation and the registration  procedures and
the probability function $\lambda_{\cal S}$.
We now add an axiom connecting this general
theoretical scheme to the usual Hilbert space  quantum
mechanics ({\bf QM} standing for  quantum mechanics).
        \begin{description}
        \item[QM] There is a bijective map $\mathcal{W}$  of
                  ${\cal K}$ onto the set $\mathcal{K}
(\mathcal{H})$ of positive self-adjoint
                  operators $W$ on a Hilbert space ${\cal H}$ with
                  ${\mbox{Tr}}(W)=1$ and a bijective map $\mathcal{F}$
                  of ${\cal L}$ onto the set $\mathcal{L}
(\mathcal{H})\subset \mathcal{B}
(\mathcal{H})$ of all self-adjoint
                  operators with $0\leq F \leq \openone$, so that
                  $\mu(w,f)=\Tr (WF)$ holds where $W=\mathcal{W}[w]$,
                  $F=\mathcal{F}[f]$.
\end{description}
Because of {\bf QM} one simply identifies ${\cal K}$ with $\mathcal{K}
(\mathcal{H})$, ${\cal L}$ with $\mathcal{L} (\mathcal{H})$ and
$\mu(w,f)$ with $\Tr (WF)$.  The convex set $\mathcal{K}
(\mathcal{H})$ is the base of the base-norm space $\mathcal{T}
(\mathcal{H})$ of trace-class operators on ${\cal H}$, while
$\mathcal{L} (\mathcal{H})$ is the order unit interval of the order
unit space $\mathcal{B} (\mathcal{H})$ of bounded operators on ${\cal
  H}$. The Banach space $\mathcal{B} (\mathcal{H})$ is the dual of the
Banach space $\mathcal{T} (\mathcal{H})$, the canonical bilinear form
being given by $\langle W,A \rangle = \Tr
(W^{\scriptscriptstyle\dagger}A)$ with $W\in \mathcal{T}
(\mathcal{H})$ and $A\in \mathcal{B} (\mathcal{H})$.  The axiom {\bf
  QM} is prepared by introducing the functions $\mu (w,f)$ and their
affine dependence on $w$. It can be guessed when quantum mechanics is
introduced in the usual textbook way and the basic statistical
interpretation is given. A deeper axiomatic effort has been done by
Ludwig\cite{Ludwig-Foundations,Ludwig-Axiomatic} in order to obtaine the
Hilbert space structure basing on physically more transparent axioms.

After the introduction of {\bf QM} we call $M$ the set of
\textsc{Microsystems}. So far we have considered only one type of
microsystems, a more general situation will be considered later on.
It seems very stimulating that the simple physical fact of essentially
statistical regularity of processes leads in Ludwig's point of view to
a concept of microsystem which goes much beyond the classical concept
of atom brought in by chemistry: it is no longer so strictly
associated with \textit{smallness} and with the role of
\textit{component} of matter.  We shall recall in the next section how
naturally mathematics of quantum theory is born if one describes this
concept of microsystem.  On the basis of the above formulation of the
foundations of quantum mechanics it is clear that the Hilbert space
does not directly describe a physical structure. It is a mathematical
tool which permits us to cleverly handle the structure of the convex
set $\mathcal{K} (\mathcal{H})$. Since the positive affine functionals
on $\mathcal{K} (\mathcal{H})$ are identical to the elements of the
positive cone of $\mathcal{B} (\mathcal{H})$ (of which $\mathcal{L}
(\mathcal{H})$ is the basis), it is the structure of $\mathcal{K}
(\mathcal{H})$ alone which determines the physical structure of
microsystems. 

\section{The modern formulation of quantum mechanics}
\label{modern}

In the introduction we have tried to give a brief exposition of the
main ideas behind Ludwig's axiomatic approach to quantum mechanics.
One of his aims was to put aside the ill-defined notions of state and
observable, primarily focusing on a proper description of the
statistical experiments one is actually faced with in quantum
mechanics. In a typical experiment a macroscopic apparatus realizing a
classically described preparation procedure triggers some detector
which gives as output a macroscopic signal, according to a suitably
devised registration procedure. The notion of microsystem is only
recovered as a convenient way to describe the most simple among such
statistical experiments, in which a preparation procedure triggers
with a definite reproducible frequency some registration procedure,
the microsystem acting as correlation carrier from the former to the
latter.  The mathematical entity describing an equivalence class of
preparation apparatuses is then identified with the state of the
microsystem, while the mathematical entity corresponding to an
equivalence class of registration apparatuses, originally called
\textit{Effekt} by Ludwig, contains the information about what has
been experimentally measured\cite{Kraus}.  The spaces in which these
objects live are the latter the dual of the former, the relative
frequency with which the preparation triggers the registration is
obtained by using the canonical bilinear form among the two spaces. This
frequency characterizes the yes-no answer of the registration or
measuring apparatus when affected by the preparation apparatus. As a
result of Ludwig's analysis in the quantum case states, to be seen as
mathematical representatives of equivalence classes of actual
preparation procedures, are given by statistical operators, while
observables, to be seen as mathematical representatives of equivalence
classes of actual registration or measuring procedures, are given by
effects. Taking into account the fact that registration apparatuses
associated to effects are naturally endowed with a scale (e.g. an
interval on the real line for a positive measurement) the notion of
effect immediately leads to the concept of observable as positive
operator-valued measure\cite{Grabowski}.  Note that the consideration
of equivalence classes is actually a key point. Utterly different and
incompatible (in the sense that they cannot be performed together)
preparation procedures might lead to one and the same state, i.e.
statistical selection procedure.  The different preparation procedures
in the same equivalence class are related to the, generally infinite,
possible decompositions of a given statistical operator, corresponding
to generally incompatible macroscopic procedures, as stressed by the
EPR paradox. In the present paragraph we will give a very brief
presentation of the more general and more flexible formulation of
quantum mechanics, which naturally comes out of Ludwig's approach.
This modern formulation of quantum mechanics, giving the most general
description of statistical experiments and transformation of states,
is obviously the result of research work by very many authors, often
starting from quite different standpoints. Among the many possible
references on the subject we recall the work by
Ludwig\cite{Ludwig-Foundations} and by Holevo\cite{HolevoOLD,HolevoNEW},
referring to these books for a more extensive bibliography. Let us
mention that the modern formulation of quantum mechanics can also be
recovered within the Bohmian approach\cite{Duerr04}.

\subsection{Description of quantum measurements}
\label{sec:stat-struct-quant}

A state in quantum mechanics, to be understood as the mathematical
representative of an equivalence class of preparation procedures, is
given by a statistical operator, i.e. a trace class operator, positive
and with trace equal to one. We recall that the set $\mathcal{T}
(\mathcal{H})$ of trace class operators on a Hilbert space
$\mathcal{H}$ form a Banach space and is in particular an
ideal of the Banach space of bounded operators $\mathcal{B}
(\mathcal{H})$, which is the dual space of $\mathcal{T}
(\mathcal{H})$, the duality form being given by the trace. In
particular the set of statistical operators $\mathcal{K}
(\mathcal{H})$
\begin{displaymath}
   \mathcal{K}
(\mathcal{H})=\{ \rho\in \mathcal{T}
(\mathcal{H}) | \rho\geq 0 \quad \Tr \rho=1 \}
\end{displaymath}
is a convex subset of the space of self-adjoint elements in $\mathcal{T}
(\mathcal{H})$ and is the base of the cone of positive elements which
generates the space of self-adjoint elements in $\mathcal{T}
(\mathcal{H})$. The convex structure of the set naturally accounts for
the possibility to consider statistical mixtures, i.e.
\begin{displaymath}
   \rho_i\in \mathcal{K}
(\mathcal{H}), \quad \lambda_i\geq 0 \quad \sum_i \lambda_i=1
\Rightarrow \sum_i \lambda_i\rho_i\in \mathcal{K}
(\mathcal{H}),
\end{displaymath}
while pure states in the sense of one dimensional projections appear as
extreme points of the convex set $\mathcal{K}
(\mathcal{H})$, i.e. elements which do not admit any further
demixture
\begin{displaymath}
   \rho=\lambda \rho_1+ (1-\lambda)\rho_2 \quad 0<\lambda<1 \quad \rho_1,\rho_2\in\mathcal{K}
(\mathcal{H}) \Rightarrow \rho=\rho_1=\rho_2,
\end{displaymath}
corresponding to the highest control in the preparation
procedure. Being compact and self-adjoint any statistical operator can
be represented as a convex combination of pure states
\begin{displaymath}
   \rho=\sum_{i} \lambda_i |\psi_i\rangle\langle \psi_i|
\quad
\lambda_i\geq 0 \quad \sum_i \lambda_i=1 \quad \|\psi_i\|=1.
\end{displaymath}
One such representation is given by the spectral representation of
$\rho$, however in general infinitely many such representations are
possible, not necessarily involving orthogonal vectors; these
different representations do generally correspond to different and
incompatible preparation procedures, in the sense that they cannot be
performed together (think e.g. of a device preparing spin 1/2
particles in terms of their spin states, fully unpolarized states can
be obtained by observing the spin along any axis, no device however
can simultaneously measure the spin along two different axes).  An
important confirmation that statistical operators give the most
general mathematical representative of a preparation comes from the
following highly nontrivial theorem by Gleason\cite{Gleason}. Let us
consider the set $\mathcal{P} (\mathcal{H})$ of orthogonal projections
in $\mathcal{H}$, in one to one correspondence with the closed
subspaces of $\mathcal{H}$, building up the so-called quantum logic of
events\cite{Cassinelli}. We first define a probability measure on
$\mathcal{P} (\mathcal{H})$ as a real function $\mu: \mathcal{P}
(\mathcal{H}) \rightarrow \mathbb{R}$ such that $0\leq \mu (E)\leq 1 \
\forall E\in \mathcal{P} (\mathcal{H})$ and $\mu (\sum_i E_i)=\sum_i
\mu (E_i)$ for $\{ E_i\}\subset \mathcal{P} (\mathcal{H}), E_iE_j=0\
i\ne j$ (i.e. $\{ E_i\}$ are compatible projections corresponding to
orthogonal subspaces).  Then according to Gleason for $dim
\mathcal{H}\geq 3$ any such probability measure has the form $\mu (E)=
\Tr \rho E$ $\forall E\in \mathcal{P} (\mathcal{H})$, with $\rho$ a
statistical operator.  

\subsection{Generalized notion of observable}
\label{sec:gener-noti-observ}

In order to describe the statistics of a given experiment, once the
state has been characterized one needs to specify the probability that
the registered value of the quantity one is trying to measure lies in
a given interval within the physically allowed range (in the following
$\mathbb{R}$ for the sake of simplicity). This amounts to define an
affine mapping (i.e. preserving convex linear combinations) from the
convex set $\mathcal{K} (\mathcal{H})$ of statistical operators to the
set of probability measures on $\mathcal{B} (\mathbb{R})$ (the Borel
$\sigma$-algebra on the space of outcomes $\mathbb{R}$). In full
generality such mappings take the form $\Tr \rho F (M)$ where $\rho$
is the statistical operator, and $F (M)$ is a uniquely defined
positive operator-valued measure\cite{HolevoNEW}. ($M$ being an
element in the Borel $\sigma$-algebra $\mathcal{B} (\mathbb{R})$). As
it is well--known a positive operator-valued measure is a mapping
defined on the $\sigma$-algebra $\mathcal{B} (\mathbb{R})$ and taking
values in the space of positive bounded operators such that $0\leq F
(M)\leq 1$, $F(\mathbb{R})=1$ so that one has the normalization
necessary for the probabilistic interpretation, and
$\sigma$-additivity holds, in the sense that $F (\cup_i M_i)=\sum_i F
(M_i)$ for any disjoint partition $\{ M_i\}$ of $\mathbb{R}$. For
fixed $M\in\mathcal{B} (\mathbb{R})$ the operator $F (M)$ is an
effect, i.e. a positive operator between 0 and $\openone$, and $\Tr
\rho F (M)$ tells us the probability that in an experiment, whose
preparation procedure is described by $\rho$, we will actually find
that our registration procedure gives a positive answer to the
question whether the measured quantity lies in the fixed interval $M$.
The above introduced structures in which registrations are naturally
associated to points or intervals in $\mathbb{R}$ is straightforward
in Ludwig's construction of effects.  Note that the statistical nature
of the experiment only requires $\Tr \rho F (M)$ to be a number
between zero and one. There is no reason to ask $F (M)$ to be a
projection-valued measure, and therefore that for any fixed
$M\in\mathcal{B} (\mathbb{R})$ the operator $F (M)$ is an orthogonal
projection (also called \textit{decision effect} by Ludwig), this will
only happen for a subset of the possible registration procedures,
corresponding to most \textit{sensitive} measurements, which moreover
do generally not exhaust the set of extreme points of the convex set
of positive operator-valued measures.  If $F (M)$ is a
projection-valued measure, it then uniquely corresponds to the
spectral measure of a self-adjoint operator in $\mathcal{H}$.  The
usual notion of observable in the sense of a self-adjoint operator is
thus recovered for particular observables. What one is really
interested in is the probability distribution of the different
possible outcomes of an experiment, once the state has been fixed, and
only in particular, though often very relevant, cases this can be done
by identifying a self-adjoint operator and its associated spectral
measure. Note that contrary to classical mechanics different
generalized observables have different probability distributions and
not all observables can be described in terms of a joint probability
distribution, leading to a structure known as quantum probability,
generalizing the classical notion of probability
theory\cite{FagnolaPROYEC-Fannes-Strocchi}. In fact it has been argued
that the passage from classical to quantum theory is actually a
generalization of probability theory\cite{StreaterJMP}.  Note that
inside Ludwig's point of view coexistence of observables is related to
the actual possibility of constructing concrete measuring apparatuses.
This (most concise) presentation of how to express the quantum
mechanical theoretical predictions for a statistical experiment,
leading to the notion of state as statistical operator and generalized
observable as positive operator-valued measure (also called
non-orthogonal resolution of identity in the mathematical literature)
is certainly not in the spirit of textbooks on quantum mechanics (even
not very recent ones), it is much closer to the presentation of
quantum mechanics one finds in the introductory chapters of books
concerned with quantum information and communication,
e.g.\cite{Nielsen-Chuang}.  Note however that in quantum information
and communication one is often only concerned with finite-dimensional
Hilbert spaces, so that the range of the positive operator-valued
measure is given by a denumerable set of operators $0\leq F_i\leq
\openone$ summing up to the identity, $\sum_i F_i = \openone$.

\subsection{Measurements as mappings on states}
\label{sec:meas-as-mapp}

Up to now we have only given the general description of the statistics
of the outcomes of a possible measurement. More generally one might be
interested in how a state is transformed as a consequence of a given
measurement.  Note that the shift from pure states, corresponding to
state vectors, to statistical operators from a mathematical standpoint
shifts the attention from operators in $\mathcal{H}$ to affine
mappings on the convex set $\mathcal{K} (\mathcal{H})$. The way in
which a state is changed as a consequence of some registration
procedure applied to it is generally described in terms of an
instrument, a notion first introduced by Davies and
Lewis\cite{Davies-Lewis}. An instrument is a mapping $\mathcal{M}$
defined on the $\sigma$-algebra $\mathcal{B} (\mathbb{R})$ giving the
possible outcomes of an experiment and taking values in the set of
operations, i.e. of contracting, positivity preserving affine mappings
on $\mathcal{K} (\mathcal{H})$, first introduced by Haag and
Kastler\cite{Haag-Kastler} and called \textit{Umpraeparierung} by
Ludwig in his axiomatic construction. In particular an instrument
$\mathcal{M}$ is such that: $\mathcal{M} (M)$ is an operation $\forall
M\in\mathcal{B} (\mathbb{R})$, i.e.  $\mathcal{M} (M)[\rho]\geq 0$ and
$\Tr \mathcal{M} (M)[\rho]\leq \openone$; $\Tr \mathcal{M}
(\mathbb{R})[\rho]=1$, accounting for normalization; $\mathcal{M}
(\cdot)$ is $\sigma$-additive, i.e.  $\mathcal{M} (\cup_i M_i)=\sum_i
\mathcal{M} (M_i)$ for any collection of pairwise disjoint sets $\{
M_i\}$. The interpretation is as follows, $\mathcal{M} (M)[\rho]$
gives the statistical subcollection obtained by selecting the prepared
state described by $\rho$ according to the fact that the measurement
outcome lies in $M$, $\mathcal{M} (\mathbb{R})[\rho]$ is the
transformed state obtained if no selection is made according to the
measurement outcome. Of course knowledge of the instrument
corresponding to a certain state transformation related to a given
measurement also provides the full statistics of the outcomes,
obtained by the positive operator-valued measure given by
$\mathcal{M}' (M)[\openone]$, where the prime denotes the adjoint with
respect to the trace operation. However different instruments actually lead to
the same positive operator-valued measure, according to the fact that
the very same quantity can be actually measured in different ways,
leading to states which transformed differently, depending on the
actual experimental apparatus used in order to implement the
measurement. Knowledge of the transformed state allows to deal with
subsequent measurements, both discrete and continuous, and in fact the
notion of instrument leads to a formulation of continual measurement
in quantum mechanics.  The field of continual measurement is by now
well established, providing the necessary theoretical background for
important experiments in quantum optics (see \cite{BarchielliLNM} for
a recent review mainly in the spirit of quantum stochastic
differential equations and\cite{Davies,continue1-continue2} for
earlier work). Once again this description of a measurement as a
repreparation of the incoming state depending on the measurement
outcome is certainly not emphasized in quantum mechanics textbooks,
but is a natural and fruitful standpoint in quantum information and
communication theory. Actually the very notion of microsystem as
something which is prepared by a macroscopic apparatus and
subsequently registered in a registration apparatus, i.e. as
correlation carrier between macroscopically operated apparatuses,
naturally emerging from Ludwig's axiomatic studies, is a very pregnant
and fertile viewpoint in quantum mechanics and in particular quantum
information and communication, as advocated by Werner\cite{Alber}; not
by chance key concepts like preparation and registration are naturally
renamed as sender and receiver.

\subsection{Open systems and irreversibility}
\label{decoh}

As a last remark we note that the operational approach we have most
briefly and incompletely sketched, stressing the relevance of mappings
acting on states living in the space of trace class operators,
corresponding to transformation of states (Schr\"odinger picture),
together with the adjoint mappings acting in the space of bounded
operators (Heisenberg picture), does not only apply to the description
of a measurement process. These mappings also generally describe the
spontaneous repreparations of a system with elapsing time, i.e. its
dynamics. If the system is closed, so that one has reversibility,
then it can be shown that the time evolution mapping necessarily has
the form $\mathcal{L}[\rho]=U (t)\rho U^{\dagger} (t)$, with $U (t)$ a
unitary mapping, and no measuring decomposition applies, consisting in
sorting statistical subcollections on the basis of a certain
measurement outcome. In the general case of an open system however
irreversibility comes in, either due to the interaction with some
environment or to the effect of some measuring apparatus, so that more
general mappings appear in order to describe this wider class of
transformations of quantum states and observables. This is an open and
very active field of research, of interest both to mathematicians and
physicists\cite{HolevoNEW,Petruccione}, where the relevance of
concepts and techniques inherited and generalized or inspired by the
classical theory of probability and stochastic processes cannot be
overstressed. A general characterization of such mappings has been
obtained only in a few cases, exploiting their property of being
completely positive.  For example in the description of irreversible
and Markovian dynamics a landmark result has been obtained by Gorini,
Kossakowski, Sudarshan and Lindblad\cite{GoriniJMP76-Lindblad},
leading to the so-called Lindblad structure of a master-equation, very
useful in applications\cite{Alicki,Petruccione}. Important hints and
restrictions on the structure of such mappings come from the
requirement of covariance under the action of some symmetry group
relevant for the system at hand\cite{HolevoNEW}. Biased by our
interests and work let us quote recent results in this framework,
dealing with quantum Brownian motion\cite{art3-art5-art6-art7-art10}
and decoherence due to momentum transfer events\cite{garda03-art12},
where the relevance of covariance and probabilistic concepts appear at
work.

\section{From microsystems to macrosystems}
\label{microtomacro}

In Ludwig's point of view space-time symmetries arise as follows: let
us consider an experiment with preparation part ${\cal Q}$ and
registration part ${\cal R}_0$ ${\cal R}$, then placing a reference
frame on ${\cal Q}$ one gets a family of symmetry transformed
registration parts $g {\cal R}_0$ $g{\cal R}$ for any reference frame
transformation $g$ belonging to the relevant symmetry group and a new
experiment ${\cal Q}$, $g {\cal R}_0$ $g{\cal R}$ can be considered. By
an appropriate treatment\cite{Ludwig-Foundations} one recovers the
typical results of usual symmetry theory based on Wigner's theorem,
where the Hilbert space $\mathcal{H}$ associated with a single
microsystem carries a unitary projective representation of the Galilei
group: if the microsystem is elementary such a representation is
irreducible\cite{Mackey}.

So far only a single microsystem has been treated, however one has
evidence of different elementary microsystems and of a huge set of non
elementary ones.  The general description of different types of
microsystems, labelled as $1,2,\ldots,n$ requires a $n$-uple of
Hilbert spaces $\mathcal{H}_1,\ldots,\mathcal{H}_n$, an element of
${\cal K}$ being a $n$-uple of positive trace class operators
$W_1,\ldots,W_n$, normalised according to $\sum_{i=1}^n \Tr W_i =1$,
an element of ${\cal L}$ being a $n$-uple $F_1,\ldots,F_n$ of
operators on $\mathcal{H}_1,\ldots,\mathcal{H}_n$ such that $0\leq F_i
\leq \openone_i$, $i=1,\ldots,n$, then finally $\mu (w,f) =
\sum_{i=1}^n \Tr (W_i F_i)$.  The projection $F_i =
(0,\ldots,0,\openone_i,0,\ldots,0)$ with probability $\mu (w,f_i) =
\Tr (W_i)$ corresponds to the registration of the microsystem of type
$i$.  It turns out that non elementary microsystems are described in
Hilbert spaces
\begin{equation}
\mathcal{H}_i = h^{(\mathrm{e})}_{\alpha_1} \otimes \ldots \otimes h^{(\mathrm{e})}_{\alpha_{\kappa_i}} 
\label{eq:mm2} 
\end{equation}
where $h^{(\mathrm{e})}_{\alpha_j}$ is the Hilbert space of an
elementary microsystem (the superscript $(\mathrm{e})$ standing for
elementary) and ${\kappa_i}$ is the number of elementary components:
the basic simplification is the restricted number of the latter ones.
A large variety of non elementary microsystems is understood having as
elementary microsystem electron and nuclei in the context of
electromagnetism; by deeper understanding it was discovered that
nuclei are not elementary microsystems and a smaller number of more
fundamental microsystems is introduced in present day subnuclear
physics. In the tensor product (\ref{eq:mm2}) many factors are
repeated: of these repeated factors only the completely symmetric or
antisymmetric part must be taken. This is a very important correction
on the simple structure (\ref{eq:mm2}), that we indicate simply by
$\mathcal{H}^{\sigma}_{i}$, the superscript $\sigma$ standing for the
aforementioned symmetrizations.

In the non relativistic case symmetry transformations for non elementary microsystems 
can be obtained from those associated with the elementary components. 
Here the self-adjoint generators of the one-parameter subgroups acquire an outstanding 
importance and the projection-valued measures associated with them provide observables 
with a straightforward physical interpretation like position, momentum, angular momentum 
and energy. Apart from the energy they have an additive structure and their expectation values 
take a simple form:
 \begin{equation}
 < A > = \sum_i \Tr_{\mathcal{H}^{\sigma}_{i}} (A_i W_i) \qquad 
 A_i = \sum_{l=1}^{\kappa_i} A^{(\mathrm{e})}_l .
 \label{eq:mm3} 
 \end{equation} 
In the case of the energy a non additive contribution generally arises
called \textit{interaction energy}, responsible of binding elementary
microsystems to composed microsystems.

When a classical limit holds the picture appears of elementary
microsystems as elementary classical particles, of composed
microsystems as structures of these interacting particles, showing
quantities built up collectively by elementary contributions: then
a composed system with very many components is a macrosystem. A phase
space $\Gamma$ emerges, a state of the macrosystem being a point $P$
in this space, a selection procedure can be represented by suitable
subsets of $\Gamma$ and it becomes a statistical selection procedure
when a probability density $\rho (P)$ is given on $\Gamma$. Of course
$\Gamma$ is a huge space, to give $\rho (P)$ and calculate the
functions $\lambda (a,b)$ can be very difficult. Actually on this
way experimental settings are invented, realized and finally work,
also a feeling is established which helps in correct guessing of $W$
and $F$ associated to $a \in {\mathcal Q}$ and $b_0 b \in {\mathcal R}_0
{\mathcal R}$.  However all this is an approximation which e.g. cannot
really grasp the typical quantum feature of $\mathcal{H}^{\sigma}_{i}$
replacing ${\mathcal{H}}_i$: it is the absence of
$\Gamma$ that makes it difficult to represent statistical selection
procedures. Then a problem appears to close in a consistent way
Ludwig's point of view inside present day quantum theory.  Ludwig aims
to a more comprehensive theory, which should provide in a natural way
a state space for a macroscopic system.
We shall now conclude this discussion indicating briefly a way we have
taken to face this problem\cite{torun99-holevo-qic}. First of all let
us stress a peculiar role that quantum field theory can have with
respect to macrosystems. A macrosystem is the physical support of all possible types of microsystems; 
we shall not base on the naive atomistic point of view that it is composed by them, instead it 
is the carrier of all of them.
If we consider the microsystems prepared when a macrosystem evolves until a time $t$, 
the $W_t \in {\mathcal K}$ shows by the structure 
$W_{1t},W_{2t},\ldots,W_{nt}$ which types of microsystems have been prepared; 
this typology varies with time $t$: the number of microsystems $N_{it}$ 
becomes an interesting quantity. The question immediately arises of an
underlying Hilbert space such that $\mathcal{H}^{\sigma}_{i}$ are isomorphic to 
subspaces of it and possibly the connection $\mathcal{H}_i \to \mathcal{H}^{\sigma}_{i}$ 
becomes natural; then an observable arises  to be interpreted as number of microsystems of type $\alpha$. 
It is well known how quantum field theory  solves in a brilliant way 
this question: for each elementary microsystem 
a Fock space ${\mathcal{H}_{F}}_{\alpha}$ is defined and the Hilbert space
is given by 
\begin{equation}
   \mathfrak{H}=\prod_{\alpha}\otimes {\mathcal{H}_{F}}_{\alpha}
\label{eq:4}   
\end{equation}
where the factors are the Fock spaces associated to each type of
elementary microsystem.
In this setting \eqref{eq:mm3} is replaced by $< A > = \Tr (A W)$,
$W$ being a statistical operator on $\mathfrak{H}$
and $A$ a self-adjoint operator in $\mathfrak{H}$.

\subsection{Quantum field theory and macrosystems}
\label{sec:from-micr-macr}

While the operators on the Hilbert spaces ${\mathcal H}_i$ are
constructed in terms of fundamental operators $x$ and $p$ having the
meaning of position and momentum with a clear classical limit, so that
quantum theory appears close to classical atomistic physics via a
\textit{quantization} procedure, in this new setting related to
$\mathfrak{H}$ given by~\eqref{eq:4} fundamental operators appear by
which $W$ and $A$ are constructed that just connect the subspaces
characterized by a fixed number of elementary microsystems, acting as
creation and annihilation operators of the elementary microsystems.
Therefore the Hilbert space $\mathfrak{H}$ and the related set of
statistical operators appear as natural candidates for a quantum
theory of a macrosystem and one can expect that just focusing quantum
field theory to macrosystems one can both improve the characterization
of physically meaningful statistical operators in $\mathfrak{H}$ and
account for the objectivity elements which should characterize
macrosystems. One is immediately confirmed in this idea by the fact
that just this new framework provides fields observables $A
(\mathbf{x})$ as densities of generators of symmetry transformations,
which obey to typical balance equations, so that $\Tr A (\mathbf{x})
W=\langle A (\mathbf{x})\rangle$ can be interpreted as expectation of
the physical quantities that one needs in the phenomenological
description of macroscopic systems. Furthermore there is the well
known example of a macrosystem at equilibrium. It is described by the
statistical operator:
\begin{equation}
   \label{eq:1}
   W\equiv \frac{e^{-\beta (H_{\Omega}-\mu N)}}{\Tr e^{-\beta (H_{\Omega}-\mu N)}},
\end{equation}
which is built in terms of the relevant observables energy $H_{\Omega}$
and the new observable $N$ typical for the passage $h^{(\mathrm{e})}_{\alpha}\rightarrow  {\mathcal{H}_{F_{\alpha}}}$ in the simplest case
of only one type of elementary microsystem. $H_{\Omega}$ is the
operator in $\mathfrak{H}$ constructed in terms of an energy density $H (\mathbf{x})$:
$H_{\Omega}=\int_{\Omega} d^3\! \mathbf{x} \, H (\mathbf{x})$, where
$H (\mathbf{x})$ is obtained in terms of the fundamental field
operator $\psi (\mathbf{x})$ as it is established by quantum field
technique, taking also in account boundary conditions on $\Omega$. The
input of all this is the Hamilton operator which comes from time
translations of a microsystem; actually by this resetting one ends up
with a self-adjoint operator $H_{\Omega}$ having a point spectrum, so
that the trace class operators $W$ can be constructed, when the
parameters $\beta$ and $\mu$ are in appropriate ranges. These
parameters label the different equilibrium macrosystems and have a
precise phenomenological meaning as temperature and chemical
potential, entering in a primary way in any macroscopic selection
procedure.

\subsection{The role of non-equilibrium states}
\label{sec:role-non-equilibrium}

The statistical operator given in~\eqref{eq:1} is an element of $\mathcal{K}
(\mathcal{H})$, constructed as a function of the observables
$H_{\Omega}$ and $N$, so that the total mass is related to a
superselection rule. The impact from thermodynamics at equilibrium is so
fruitful that one wonders whether one can generalize it outside the
very particular and in a sense too strongly idealized situation
described as \textit{equilibrium}, still satisfying the superselection
rule for the total mass. Once a suitable set of
\textit{relevant} linearly independent field observables $A_j
(\mathbf{x})$ is given in $\mathfrak{H}$ one considers a set of
classical fields $\zeta_j (\mathbf{x})$ such that the operator $\Phi
(\zeta)\equiv \sum_j\int_{\Omega} d^3\! \mathbf{x}\, \zeta_j
(\mathbf{x})A_j (\mathbf{x})$ is essentially self-adjoint and
$e^{-\Phi (\zeta)}$
is trace class so that a
statistical operator, that we call macroscopic reference state, can be
defined:
\begin{equation}
 \label{gibbsref}
W_{\zeta}= \frac{\mathrm{e}^{-\Phi (\zeta)}}{\Tr \mathrm{e}^{-\Phi (\zeta)}}.
\end{equation}
The classical fields represent a \textit{local generalization} of the previous
equilibrium parameters $\beta$, $\mu$: the field operators $A_j
(\mathbf{x})$ have a quasi-local character in the sense that they
depend on ${\psi}(\mathbf{x})$,
${\psi}^{\dagger} (\mathbf{y})$ for
$|\mathbf{x}-\mathbf{y}|\ll\delta$ with $\delta$ much smaller than the
typical variation scale of the state parameters $\zeta_j
(\mathbf{x})$. Such a quasi-local character emerges if one considers
the fundamental mechanical densities, that we recall in the
non-relativistic case: mass density, momentum density, kinetic energy
density, where higher derivatives inside the different expressions
loosely mean less locality. The field $\psi(\mathbf{x})=\int d^3\!
\mathbf{x}_1 \ldots d^3\! \mathbf{x}_k\, g
(\mathbf{x},\mathbf{x}_1,\ldots,\mathbf{x}_k) \psi_1
(\mathbf{x}_1)\ldots\psi_k (\mathbf{x}_k)$ refers to a field
composed by elementary ones, $g
(\mathbf{x},\mathbf{x}_1,\ldots,\mathbf{x}_k)$ being a suitable
structure function of microsystems, concentrated for
$|\mathbf{x}-\mathbf{x}_{i}|\ll\delta$, $i=1,2\ldots,k$.  The
reference state~\eqref{gibbsref}, that we call a \textit{macrostate}
with state parameters $\zeta(\mathbf{x})$, provides a geometrical
structure\cite{Streater} which replaces lacking phase--space and the
expression $-k\Tr (W_{\zeta}\log W_{\zeta} )$ acquires the role
of thermodynamical entropy.  The subtlety with relevant variables is
that their linear span is not invariant under time evolution.
The far reaching consequence of this is that the general dynamics of a
macrosystem cannot be described only with a family of
\textit{macrostates} $W_{\zeta_t}$ by a suitable choice of time
dependent state parameters ${\zeta_t}$: a more general framework is
necessary and in addition to relevant variables, \textit{irrelevant}
ones impose themselves.  One succeeds however in constructing
statistical operators $\rho_t$, solutions of the Liouville-von Neumann
equation, having as input the reference state~\eqref{gibbsref} and
displaying the whole history $\zeta_{t'}(\mathbf{x})$ for $t'<t$ of
the state parameters and also introducing irreversibility in a
fundamental way.  This is also the philosophy behind the formalism of
non-equilibrium statistical operator initially proposed by
Zubarev\cite{Zubarev} and extensively used in non-equilibrium
thermodynamics\cite{Roepke}. Now we come to the main difficulty: the
construction of $\rho_t$ by means of \textit{one} state
$W_{\zeta_{t}}$, i.e. $\rho_t$ carrying only \textit{one} family of
state parameters, is in general successful only for suitable time
intervals. Mixtures of several reference states with the
structure~\eqref{gibbsref} but modified through the creation of
microsystems in suitable states ${\psi_\alpha}_t\in\mathcal{H}_i$ also
appear, providing a much richer parametrization of $\rho_t$ in terms
of states ${\psi_\alpha}_t$ of microsystems with certain statistical
weights and state parameters ${\zeta_{\alpha}}_t$ influenced by the
microsystem. In conclusion on the space $\mathfrak{H}$ state
parameters, having a direct relevance in macroscopic phenomenology,
can be introduced in a natural way. Deterministic evolution of one set
of state parameters means that a selection has been done fine enough
to avoid for them a statistical description; in general however, at
least piecewise, dynamics is not deterministic and this is described
by the appearance of microsystems.

Looking at quantum field theory in this way, microsystems become
intertwined with the reference macrostate, which in a sense replaces
the vacuum state of the usual field theoretical treatment of
microsystems: this can have an impact also for the general description
of interaction in the context of relativistic quantum field theory.

\section{Conclusions and Outlook}
\label{sec:conclusions-outlook}
We have recalled the modern formulation of quantum theory with
observables given by positive operator-valued measures and evolution
of possibly open systems given by mappings on states, recalling in
particular the starting point of Ludwig's approach. In this framework
the essentially statistical character of the phenomenology of
microsystems appears as a universal feature of typical non equilibrium
systems. We have argued that quantum field theory, which emerges as
underlying theory of all microsystems, appears as a natural framework
in which statistical operators can be constructed carrying objective
state parameters consisting in classical fields, which generalize the
well known parametrization in terms of temperature and chemical
potential. Evidence of microsystems is related with the breakdown of
the deterministic time evolution of these state parameters: then by
facing stochasticity in their dynamics, quantum theory of microsystems
can emerge in a natural way from quantum field theory, initially
focused on macrosystems, putting in a new light the question of a
proper separation of their dynamics from the macroscopic background.

\begin{acknowledgments}
   The authors gratefully acknowledge financial support by MIUR under FIRB
   and PRIN05.
\end{acknowledgments}

\end{document}